\documentclass[pre,aps,floatfix,twocolumn,showpacs,preprintnumbers,superscriptaddress]{revtex4-1}
\usepackage{graphicx}
\usepackage{dcolumn}
\usepackage{bm}
\usepackage{amssymb}
\usepackage{pifont}
\usepackage{amsmath}
\usepackage{times}
\usepackage{color}

\begin{document}

\title{Self-sustained irregular activity in an ensemble of neural oscillators}

\author{Ekkehard Ullner}
\email{e.ullner@abdn.ac.uk}
\affiliation{Institute for Complex Systems and Mathematical Biology and SUPA,
University of Aberdeen, Aberdeen AB24 3UE, United Kingdom}

\author{Antonio Politi}
\email{a.politi@abdn.ac.uk}
\affiliation{Institute for Complex Systems and Mathematical Biology and SUPA,
University of Aberdeen, Aberdeen AB24 3UE, United Kingdom}

\date{\today}
\begin{abstract}
An ensemble of pulse-coupled phase-oscillators is thoroughly analysed in the presence of 
a mean-field coupling and a dispersion of their natural frequencies. In spite of the
analogies with the Kuramoto setup, a much richer scenario is observed.
The ``synchronized" phase, which emerges upon increasing the coupling strength, is
characterized by highly-irregular fluctuations: a time-series analysis reveals
that the dynamics of the order parameter is indeed high-dimensional.
The complex dynamics appears to be the result of the non-perturbative action
of a suitably shaped phase-response curve. Such mechanism differs from the often invoked 
balance between excitation and inhibition and might provide an alternative basis 
to account for the self-sustained brain activity in the resting state.
The potential interest of this dynamical regime is further strengthened by its (microscopic) linear
stability, which makes it quite suited for computational tasks. 
The overall study has been performed by combining analytical and numerical studies, 
starting from the linear stability analysis of the asynchronous regime, to include the Fourier
analysis of the Kuramoto order parameter, the computation of various types of Lyapunov exponents, 
and a microscopic study of the inter-spike intervals.

\end{abstract}


\maketitle

\section{Introduction} 

Most of the challenging questions that arise in the attempt of improving our understanding 
of the natural (and artificial) world deal with multi-component systems, whose overall dynamics
is the result of many nonlinear interactions.
The difficulty of the task is often mitigated by the assumption of being before
universal phenomena, which do not crucially depend on the details of the
underlying models. It is therefore customary to deal with relatively simple setups
in the hope that relevant details are not missed.

The mammalian brain is the most prominent example where this approach is
absolutely necessary, if we wish to make some substantial progress.
There, even after disregarding several ingredients, such as 
the multiple degrees of freedom involved in the dynamics of realistic neurons
(as in multicompartmental models~\cite{Gerstner-Kistler-02}), the diversity among the 
single units, the topology and the plasticity of the connections, the range of
possible dynamical phenomena is still very rich and not yet entirely understood.
Self-consistent partial synchronization is a simple but enlightening example.
The phenomenon, discovered by van Vreeswijk in an ensemble of leaky integrate-and-fire neurons 
(LIF)~\cite{vanVreeswijk-96}, was believed for a long time to be a non-perturbative
effect. Only, recently it has been however clarified~\cite{Politi-Rosenblum-15} that it is equivalent to the
rotating waves observed in the weak-coupling limit~\cite{Hansel-Mato-Meunier-95},
and can indeed be observed and characterized in Kuramoto-Daido oscillators~\cite{Daido-92a,*Daido-93a,*Daido-96},
as well.

In general, the problem of characterizing the collective dynamics of an ensemble of oscillators
is deeply connected to the question of how different levels of descriptions are linked to one another. 
In computational neuroscience, it is customary to consider subpopulations of neurons, 
under the assumption that the firing rate is the single relevant variable, as in the
seminal paper by Wilson and Cowan~\cite{Wilson-Cowan-72} and in several other publications 
(see e.g. ~\cite{Gerstner-Kistler-02,Coombes-05,Ermentrout-Terman-10}).
It is not, however, clear whether such models can be derived starting from more
microscopic setups based on single spiking neurons.
Some recent studies have shown that a low-dimensional collective dynamics may emerge in
networks of theta (or, equivalently, quadratic-integrate-and-fire, QIF) neurons~\cite{Luke-etal-13,So-etal-14}.
More than that, a reformulation of  pulse-coupled oscillators in terms of firing-rate models
has been accomplished in~\cite{Laing-14,Montbrio-Pazo-15}.
The validity of these results is due to the existence of relationships such as the Ott-Antonsen 
Ansatz~\cite{Ott-Antonsen-08} and the Watanabe-Strogatz theorem~\cite{Watanabe-Strogatz-94}, which allow 
expressing the collective behavior in terms of a few variables, the others being essentially slaved. 
Such theoretical pillars are however based on strong simplifying assumptions on the nature
of the inter-oscillator coupling~\cite{Watanabe-Strogatz-94,Ott-Antonsen-08}.

To what extent is the compression of degrees of freedom effective in more general setups?
The background activity of the brain in the resting state, when no specific task is 
performed~\cite{Arieli-95,Tsodyks-99, Destexhe-11} testifies to a collective irregular dynamics.
Moreover, the ongoing discussion about rate- versus temporal-coding~
\cite{Shadlen-Newsome-jneurosci-98,Butts-etal-nature-07}, suggests that the firing rate
may not be sufficient to ensure the necessary computational capability of the mammalian brain.

Altogether, one should thus expect an irregular collective behavior.
It is often conjectured that the self-sustained activity
is the result of a balance between activation and inhibition~\cite{vanVreeswijk-Sompolinsky-96,vanVreeswijk-98}.
Mathematically, this means that the effect of the coupling is zero on average so that
it is essentially controlled by stochastic/chaotic fluctuations.
It is not, however, clear how such a balance can be durably ensured in self-organized
networks of firing oscillators.
A conceptually different possibility to account for a macroscopic irregularity
is offered by the nonlinear character of the Liouville-type equation (which, strictly
speaking, applies to an ensemble of infinitely many oscillators).
This functional equation operates in an infinite-dimensional
phase space and can, thereby, generate a dynamics of arbitrary complexity.
This has been indeed observed in an abstract model of coupled maps~\cite{Shibata-etal-99} 
and in globally coupled Stuart-Landau oscillators~\cite{Takeuchi-Chate-13}.
In both cases, the single dynamical units are intrinsically more complex than 
phase oscillators: the logistic maps are chaotic by themselves, Stuart-Landau
oscillators can behave chaotically under the action of a periodic modulation.
In the case of phase oscillators, there is only a preliminary evidence 
in an ensemble of LIF neurons with delayed interactions~\cite{Luccioli-Politi-10}.

In this paper, we present a model whose overall activity is intrinsically 
highly-dimensional. As this dynamical phase is rather robust against variations
of several parameters, it may provide an alternative mechanism for
the self-sustainment of the resting brain activity.
More precisely, we study an ensemble of pulse-coupled phase oscillators, whose
phase-response curve (PRC) is derived by smoothing the PRC of LIF neurons. 
Other than that, our setup is the same as in the standard Kuramoto 
model~\cite{Kuramoto-84,Acebron-etal-05}:
the single oscillators are characterized by a distribution of bare frequencies, while
the coupling is homogeneously all-to-all. As in the Kuramoto model, a synchronization
transition is observed upon increasing the coupling strength, but the analogies
end here, since above criticality, the order parameter, rather than being constant,
exhibits complex high-dimensional oscillations. 
Such fluctuations are present in the ``activity" field as well, a variable akin to the 
electric potential recorded while measuring EEGs.

As briefly discussed in section~\ref{sec:theory}, in the weak-coupling limit (and for a
small dispersion of the frequencies), our system reduces to a Kuramoto-Daido model, 
with the coupling function being composed of several Fourier harmonics.
Recent studies of such a type of models have revealed quite a rich
phenomenology (see, e.g.~\cite{Komarov-Pikovsky-14,Ashwin-Burylko-15}).
This is not, however, sufficient to account for the qualitative differences
reported in this paper: direct simulations show that the scenario hereafter 
discussed disappears when the dispersion of frequencies is decreased.

More specifically, the overall dynamics is characterized by a spectrum 
of negative Lyapunov exponents. This {\it inconsistency} is nothing
but a manifestation of 
stable chaos~\cite{Politi-Torcini-10}, an irregular dynamics of 
cellular-automaton type, which is self-sustained because of the high 
(infinite) dimensionality of the phase space (in other words, it dies
out in finite ensembles). In neural systems, stable chaos was first 
found in a diluted network of LIF units~\cite{Zillmer-etal-06}, and
later discussed in more disordered setups~\cite{Jahnke-etal-08,Zillmer-etal-09}.
At variance with deterministic chaos, accompanied by an exponential
separation of orbits and thereby a loss of memory, stable
chaos is identified by a ``microscopically" stable dynamics, which is
definitely more appropriate for the performance of computational tasks.
The potentiality of stable chaos for information processing has
been preliminarily explored in ~\cite{Monteforte-Wolf-10,Monteforte-Wolf-12}.
The onset of a macroscopic irregular dynamics, as discussed in this
paper, makes this perspective even more intriguing, for the
richness of the collective behavior.

In section \ref{sec:model} we introduce the setup and justify its choice. 
In the following section~\ref{sec:theory}
we reconstruct the asynchronous state and investigate its stability properties.
Differences and analogies with the standard Kuramoto model are emphasized.
In particular, we find that the asynchronous regime loses its stability when 
a complex eigenvalue is born out of the line containing the continuous spectrum. 
In the last part of the section, the proper order parameters for the
characterization of the transition are introduced: they are the Kuramoto order parameter,
whose definition requires passing to more appropriate phase variables, and the
activity field.
Sections \ref{sec:macro} and \ref{sec:micro} are devoted to a careful numerical
analysis of the synchronized phase at the collective and microscopic level, respectively.
Due to the difficulty of dealing with finite-size corrections, we study the resulting
behavior sufficiently far from the transition. In section \ref{sec:macro} we 
first illustrate the phase diagram and the initial part of the Lyapunov
spectrum. We then show the power spectrum of the order parameter and carry on a 
time-series analysis to determine the fractal dimension.
In section \ref{sec:micro} we focus our interest on the behavior of the single
neurons, computing the effective frequency and the conditional Lyapunov exponents:
they are all negative, indicating that we are in the presence of generalized synchronization.
The presence of phase slips is also unveiled. Finally in the last section we summarize
the main results and discuss the several perspectives that are opened by the scenario
discussed in the paper.

\section{The model}
\label{sec:model}

The starting point of this paper is the model of delayed LIF neurons studied in 
\cite{Luccioli-Politi-10}. Here, the model is modified to make it simpler, more
generic and more amenable to both numerical and analytical studies.

In this perspective, we consider an ensemble of pulse-coupled phase oscillators,
in the presence of $\delta$-like pulse and charactereized by a suitable PRC $\Gamma(\phi)$,
\begin{equation}
\dot \phi_i = \omega_i - \frac{g}{N} \Gamma(\phi_i) \sum_j \delta(t-t_j) \; ,
\label{eq:pulsecoup}
\end{equation}
where $\phi_i \in[0,1]$ is the local phase, $\omega_i$ the bare oscillator frequency 
(i.e. in the absence of coupling),
$g$ the coupling strength and $N$ is the system size.
Whenever any oscillator reaches the threshold $\phi_i=1$, a $\delta$-spike is sent and received 
by all neurons. 
The above formulation is quite general for two reasons: (i) any model where the velocity 
field $\dot \phi_i$ is phase dependent (in the absence of coupling), can be always 
rephrased as Eq.~(\ref{eq:pulsecoup}) upon suitably changing variables~\cite{Abbott-vanVreeswijk-93};
(ii) finite-width pulses can be mapped onto $\delta$-like ones, upon suitably adjusting
the shape of the PRC~\cite{Politi-Rosenblum-15} (at least in the weak coupling limit).

An important ingredient of the model studied in \cite{Luccioli-Politi-10} is the presence of a 
delay between spike emission and reception. In the weak-coupling limit, when
the dynamics is nearly homogeneous, one can simulate the presence of a delay as a suitable
phase shift of the PRC and this is what has been assumed here.
The phase shift should be different for the different
oscillators. However, here, for the sake of simplicity we assume the same PRC for all the
oscillators. 

\begin{figure}[htb]
\begin{center}
\includegraphics[width=0.9\columnwidth,clip=true]{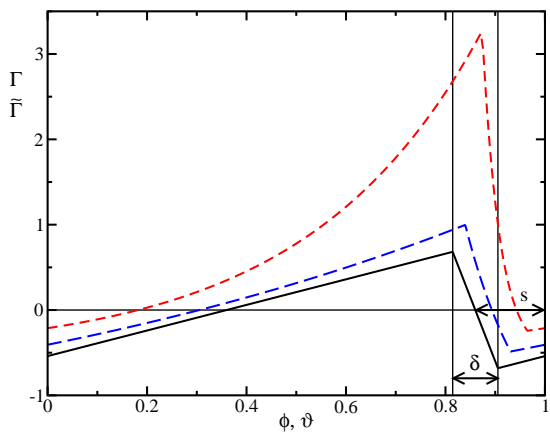}
\caption{Phase response curve $\Gamma(\phi_i)$ according to Eq.~(\ref{eq:PRC}) for the 
standard parameter values $b_1=1.5$, $s=0.14$, and $\delta=0.1$
(solid line); the short- and long-dashed curves correspond to the effective PRC 
$\tilde \Gamma$ obtained for $\omega=\omega_{min}=0.8$ and $\omega=\omega_{max}=2$, respectively
(with $g=0.8$).}
\label{fig:PRC}
\end{center}
\end{figure}

In the LIF model, $\Gamma(\phi_i) = a\exp(b\phi_i)$ where $\phi_i$ is assumed to be taken
modulus $1$. As a result of a phase shift, the discontinuity originally present when passing
from $1$ to $0$, moves inside the unit interval. For the sake of generality and simplicity,
we prefer to remove the discontinuity, considering a piece-wise linear PRC, such as
\begin{equation}
\Gamma(\phi) =
\begin{cases}
B_{01} + b_1 \phi &\mbox{if } \quad 0\le\phi<\phi_l \\
B_{02} - b_2 \phi & \mbox{if } \quad \phi_l \le \phi \le \phi_r \\
B_{03} + b_1 \phi & \mbox{if } \quad \phi_r < \phi < 1 \; ,
\end{cases}
\label{eq:PRC}
\end{equation}
where the various parameters are chosen so as to ensure continuity in $\phi_l$, $\phi_r$ and equality 
between $\phi=0$ and $1$. 
Considering that the amplitude of the PRC is controlled by the coupling constant $g$,
there are three truly independent parameters: one controlling the vertical shift of the PRC,
and two which identify the junction points. As for the first parameter, it basically controls
whether the coupling has an average excitatory or inhibitory effect, thereby inducing
a speeding up or slowing down of the spiking activity. Since we are not interested in such
effects, but rather in the mutual attraction or repulsion among the oscillators, we have decided
to assume that the PRC has zero average. 
The two remaining parameters are identified by the phase
shift $s$ (defined as the distance of the midpoint of the central region from $1$ - see the Fig.~\ref{fig:PRC}) and
the width $\delta$ of the central interval. 
Altogether, $b_2 = b_1/\delta$,
$B_{01}= b_1(s-1/2)$, $B_{02}= b_1(1-s)/\delta$,
$B_{03}= b_1(s-3/2)$, while $\phi_l = (1 - s + \delta/2 - \delta s)/(\delta + 1)$ 
and $\phi_r = (1 - s + 3\delta/2 - \delta s)/(\delta + 1)$. 
The parameter $b_1$ has been set equal to 1.5 (in principle, it can be absorbed in
the definition of $g$), while the two other parameters have been set $s=0.14$, $\delta = 0.1$
in all of the following simulations.  The resulting shape of the PRC is presented in
Fig.~\ref{fig:PRC}.

Finally, we have chosen to work with a uniform distribution of frequencies centered in 
$\bar \omega =1.4$. The simulations reported in
this paper refer to a half width $\Delta= (\omega_{max}-\omega_{min})/2 = 0.6$, but similar results have
been obtained for different values of $\Delta$. 
We evolve the model Eq.~(\ref{eq:pulsecoup},\ref{eq:PRC}) as an event driven process. 
Between two consecutive $\delta$-spikes, the phase of each oscillator increases linearly according to 
its individual bare frequency $\omega_i$. 
When one of the oscillators reaches the firing threshold $\phi_i=1$, its phase is reset to zero and 
all phases are adjusted to account for the received spike. 
The effect of the coupling might bring a second oscillator beyond the firing threshold. 
In such a case, also that oscillator is reset to zero plus an offset due to the spike received from 
the first oscillator. We continue this evolution, without advancing the time, until no further spikes 
are triggered. In practice, ``avalanches" may occur: we have controlled that they do not contribute 
significantly to the global behavior as their size does not increase upon increasing the number of neurons. 
Notice also that in the original model \cite{Luccioli-Politi-10}, avalanches do not exist.

\section{Theory}
\label{sec:theory}
Some insight can be gained by considering the thermodynamic limit, as this allows determining
analytically the properties of the stationary asynchronous regime.

First of all it is convenient to define the activity field $E(t)$ as the number of
spikes emitted per unit time, so that Eq.~(\ref{eq:pulsecoup}) can be rewritten as
(for the sake of simplicity we drop the subindex $i$),
\begin{equation}
\dot \phi = \omega - g  \Gamma(\phi) E(t) \; .
\label{eq:phase1}
\end{equation}
Let us now introduce the probability density $Q(\phi,\omega,t)$, as the fraction of
neurons with a bare frequency in $[\omega,\omega+d\omega)$, whose phase belongs
to $[\phi,\phi+d\phi)$ at time $t$. Clearly,
\[
\int Q(\phi,\omega,t) d\phi = P(\omega) \; ,
\]
where $P(\omega)$ is the density of neurons with bare frequency $\omega$.
$Q$ satisfies the continuity equation
\begin{equation}
\frac{\partial Q}{\partial t} = -\frac{\partial}{\partial \phi} \left [
\omega - g \Gamma(\phi) E(t) \right ] Q \; ,
\label{eq:continuity}
\end{equation}
while the field $E$ satisfies the self-consistent equation
\begin{equation}
E(t) = \int Q(1,\omega,t)[\omega - g\Gamma(1)E(t)]d\omega \; ,
\label{eq:field1}
\end{equation}
which implies 
\[
E(t) =  \frac{\int \omega Q(1,\omega,t)d \omega}
{1+g\Gamma(1)\int Q(1,\omega,t)d \omega} \; .
\]
The asynchronous regime corresponds to the stationary solution whose phase-dependence is 
determined by setting the time derivative of $Q$ equal to zero. 
By properly renormalizing the flux, one obtains
\begin{equation}
Q_0(\phi,\omega) = \frac{P(\omega)}{T(1,\omega)[\omega-g\Gamma(\phi) E_0]} \; ,
\label{eq:R0}
\end{equation}
where 
\begin{equation}
T(\psi,\omega,E_0) = \int_0^\psi \frac{d\phi}{\omega - g\Gamma(\phi)E_0} \equiv
\int_0^\psi d\phi \; \tau(\phi,\omega)
\label{eq:Tomega}
\end{equation}
is the time required by an oscillator with frequency $\omega$ to reach the phase $\psi$, starting from 0,
in the presence of a constant field $E_0$. $T(1,\omega,E_0)$ is 
thereby the interspike interval, while $\tau(\phi_i,\omega)$ is the inverse instantaneous effective frequency.
The field $E_0$ can be finally obtained from Eqs.~(\ref{eq:field1},\ref{eq:R0})
\begin{equation}
E_0 = \int \frac{P(\omega)}{T(1,\omega)}d \omega \; .
\label{eq:E0}
\end{equation}

The above calculation yields the structure of the asynchronous state, but it does not tell us
whether it is stable.
The stability can be assessed by investigating the behavior of infinitesimal perturbations.
Let us define,
\[
Q(\phi,\omega,t) = Q_0(\phi,\omega) + q(\phi,\omega,t) \quad , \quad  E(t) = E_0 + e(t)
\]
$q$ and $e$ satisfy the following equations,
\[
\frac{\partial q}{\partial t} = -\frac{\partial}{\partial \phi} \left [
\omega - g \Gamma(\phi) E_0 \right ] q + g e(t)
\frac{\partial \Gamma(\phi)Q_0}{\partial \phi} 
\]
and
\[
e(t) =  \frac{\int (\omega-g\Gamma(1)E_0) q(1,\omega,t)d \omega}
{1+g\Gamma(1)\int Q_0(1,\omega)d \omega} \; .
\]
These two equations can be solved by introducing a standard Ansatz,
$q(\phi,\omega,t) = u(\phi,\omega) \textrm{e}^{\mu t}$, $e(t) = z \textrm{e}^{\mu t}$.
One obtains
\begin{equation}
\mu u = g\Gamma' E_0 u -\left [
\omega - g \Gamma E_0 \right ] u'+ g \Gamma'Q_0 z
+ g \Gamma Q'_0 z
\label{eq:afterA}
\end{equation}
and
\begin{equation}
z =  \frac{\int (\omega-g\Gamma(1)E_0) u(1,\omega)d \omega}
{1+g\Gamma(1)\int Q_0(1,\omega)d \omega} \; ,
\label{eq:epsdef}
\end{equation}
where the prime denotes a derivative with respect to $\phi$ and we have dropped the depedence on
$\phi$ for the sake of simplicity.
The solution of such equation, reported in the appendix A, yields the eigenvalue equation (\ref{eq:mudef}).

The spectrum of the linear operator consists of a continuous and a discrete component.
The continuous part is confined to an interval along the imaginary axis and is therefore composed
of marginally stable directions.
The discrete component can be obtained by assuming $\mu=\mu_R+i\mu_I$, separating (\ref{eq:mudef})
into real and imaginary parts, and finally looking for the zeros in the complex plane.

\begin{figure}[htb]
\begin{center}
\includegraphics[width=0.9\columnwidth,clip=true]{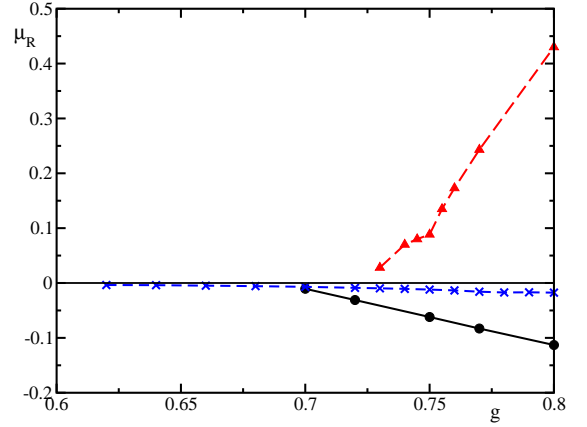}
\caption{Stability diagram of the asynchronous state. The real part of the discrete eigenvalues
is reported versus the coupling strength $g$.}
\label{fig:stability}
\end{center}
\end{figure}

A numerical study reveals the presence of (at least) three pairs of complex-conjugate eigenvalues
(see Fig.~\ref{fig:stability}, where the imaginary part is not reported) 
two of which being negative and one positive. Two pairs of exponents arise definitely above some 
finite $g$ value; the third one is likely to follow the same scenario, 
but given its small real part we could not trace it for small coupling strengths
(see the crosses in Fig.~\ref{fig:stability}).
Altogether, the asynchronous solution is marginally stable up until $g_c \approx 0.72$,
when it destabilizes for the onset of a pair of complex conjugate eigenvalues with a positive
real part. 

Let us now compare with the stability of the asynchronous solution in the Kuramoto model. 
Below criticality, in both cases the probability distribution is marginally stable (see
~\cite{Strogatz-Mirollo-91} for the first such analysis in the Kuramoto setup), the major
difference being the presence of a discrete stable spectral component in our model.
Noteworthy, in spite of the marginal stability of the probability density, the order parameter 
(see the next section for its definition) relaxes exponentially in the Kuramoto model.
This is a manifestation of the so-called Landau damping~\cite{Strogatz-etal-92}. Only recently
this ``inconsistency" has been fully resolved, by understanding that different classes of
functions may be considered in the stability analysis~\cite{Chiba-15,Fernandez-etal-14,Dietert-15}.
We do not know how much of such studies carry over to the present setup: this is an open problem.

At criticality, a pair of complex eigenvalues with a positive real part is born: this is at variance
with the standard Kuramoto model, where the newly appearing eigenvalue is real. There is, instead, an analogy with
the Kuramoto model with delay~\cite{Yeung-Strogatz-99,Choi-etal-00}, where periodic oscillations
arise. Here, however, above threshold, the probability density rather than oscillating periodically, 
behaves irregularly, as discussed in the following sections.

\subsection{Order parameters}

\begin{figure}[htb]
\begin{center}
\includegraphics[width=0.6\columnwidth,clip=true]{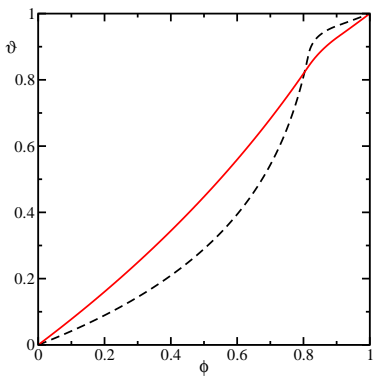}
\caption{$\vartheta(\phi)$ for $g=0.8$ and two different frequencies: $\omega=0.8$ (dashed curve)
and $\omega=2$ (solid curve).}
\label{fig:thetaphi}
\end{center}
\end{figure}

In order to study the transition, it is necessary to identify a suitable order parameter.
One cannot directly use the phase $\phi$ to define the Kuramoto order parameter~\cite{Kuramoto-84}, since
in the asynchronous regime, such phases do not advance homogeneously in time, i.e. they
are not proper phases. This can be accomplished by introducing the new variable $\vartheta$
\begin{equation}
\frac{d \vartheta}{d \phi} = \frac{\tau(\phi,\omega)}{T(1,\omega,E_0)}
\label{eq:thetaphi}
\end{equation}
that is basically equivalent to the elapsed time (apart from a scaling factor) and thus
advances uniformly by definition. With reference to this new phase, the local dynamics 
is described by the equation
\begin{equation}
\dot \vartheta = \tilde \omega - g \tilde \Gamma(\vartheta) (E(t)-E_0) \; ,
\label{eq:phase2}
\end{equation}
where $\tilde \Gamma(\vartheta)$ is the effective PRC
\begin{equation}
\tilde \Gamma(\vartheta)  =  \frac{\tilde \omega \Gamma(\phi(\vartheta))}
{\omega-gE_0\Gamma(\phi(\vartheta))} \; ,
\label{eq:effPRC}
\end{equation}
$\vartheta(\phi)$ is obtained by solving Eq.~(\ref{eq:thetaphi}), and 
$\tilde \omega = 1 / T(1,\omega,E_0)$, is the effective frequency. As it is
understood from its definition $\vartheta(\phi)$, depends both on $g$ and the bare
frequency. The dependence of $\vartheta$ on $\phi$ is reported in Fig.~\ref{fig:thetaphi}
for the maximal and minimal frequencies at $g=0.8$. The particular transformation from the phase $\phi$ 
to the new effective phase $\vartheta$ is shown in the appendix B.

Once a proper phase $\vartheta$ has been identified, a meaningful Kuramoto order parameter can be defined,
\[
R = \frac{1}{N} \sum_j \mathrm{e}^{2\pi i \vartheta_j} \; .
\]
Since $R=0$ in the asynchronous regime, it can be safely used to identify the onset of even weak forms
of synchronization. 

Given the large amount of transformations needed to determine $R$, and because of
the relationship with neural networks, we have often considered a second order parameter,
a smoothened version $Y(t)$ of the field $E(t)$. In a finite system, $E(t)$ is just
a collection of $\delta$-pulses. It is therefore more convenient to investigate
\[
\dot Y = -\gamma Y + E(t)  \; .
\]
We have selected $\gamma=5$.
In the asynchronous regime, the activity is constant, i.e. $Y_0 = E_0/\gamma$.
Above the transition that we are going to discuss, the activity starts oscillating
in time, so that it is convenient to introduce the temporal standard deviation
\[
\sigma_Y = \sqrt{\langle Y^2 \rangle-\langle Y\rangle^2} \; ,
\]
where the angular brackets denote a time average. It will be also useful to look at
the fluctuation $\sigma_R$ of the Kuramoto order parameter $R$, as this indicator allows
identifying the regimes where the degree of synchronizarion oscillates in time.

We end this theoretical section by briefly commenting on the weak-coupling, low-disorder limit.
The effective PRC $\tilde \Gamma$ depends on the frequency of the oscillator (and on the coupling
strength). The resulting curves 
for $\omega_{min}$ and $\omega_{max}$ (and $g= 0.8$) are reported in Fig.~\ref{fig:PRC}: see the dashed lines. 
In the small-disorder limit, one can neglect such a dependence. By then following, Ref.~\cite{Politi-Rosenblum-15},
we expect the model to become equivalent to the Kuramoto-Daido model
\[
\vartheta_i = \omega_i - \frac{g}{N}\sum_j \tilde \Gamma(\vartheta_i-\vartheta_j) \; .
\]
The above equation makes one difference with the Kuramoto model transparent: the sinusoidal
coupling function is replaced by the more structured function $\tilde{\Gamma}$  (see Fig.~\ref{fig:PRC}).
This is not, however, the major source of differences, since the simulations show that the 
weak-disorder assumption is not appropriate to reproduce the scenario discussed in this paper.

\section{Macroscopic dynamics}
\label{sec:macro}

\begin{figure}[htb]
\begin{center}
\includegraphics[width=1.0\columnwidth,clip=true]{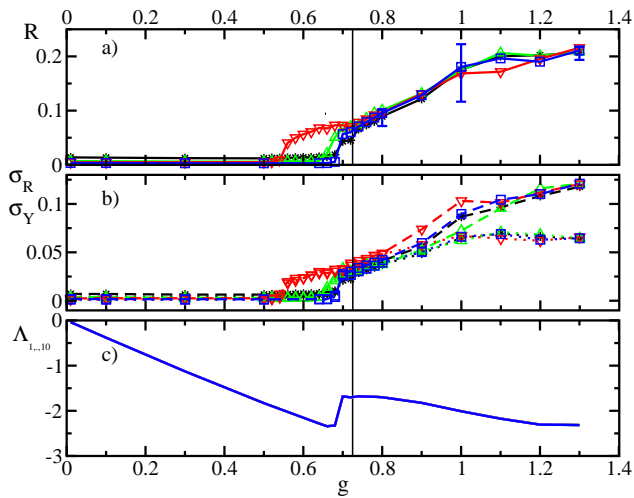}
\caption{Phase diagram: Dependence of the Kuramoto order parameter R (a), its standard deviation $\sigma_R$, 
the standard deviation $\sigma_Y$ of the activity field $Y$ (b), and the ten largest Lyapunov exponents 
versus the coupling strength $g$. 
The curves in a and b have been obtained for $N=4000$ (black), $N=16000$ (green) and $N=64000$ (blue), upon increasing
the coupling $g$. 
The red curve (a and b) is a continuation by decreasing $g$ with $N=64000$. 
$\sigma_R$ and $\sigma_Y$ are represented with dashed and dotted lines, respectively (b). 
The vertical lines mark the critical point $g_c \approx 0.72$, where the linear stability of the 
asynchronous state is lost (see section \ref{sec:theory}). 
The ten largest global Lyapunov exponents in panel c are almost indistinguishable. 
The simulations have been performed for $N=64000$, increasing $g$.}
\label{fig:phasediag}
\end{center}
\end{figure}

The most appropriate control parameter to study the onset of collective dynamics is the coupling strength $g$.
In Fig.~\ref{fig:phasediag}a,b, it is used to parametrize the dependence of the Kuramoto order parameter $R$, 
its temporal standard deviation $\sigma_R$, and the standard deviation of the activity field $Y$.

Each data point is based on a simulation over $500$ time units after a transient of $50$ time units. 
All, but the red curve, have been obtained by increasing the coupling strength $g$ stepwise, using 
the final condition for a given $g$ value as the initial condition for the next one.
The statical uncertainty (represented by the error bars) has been estimated by dividing the 
standard deviation of the Kuramoto order parameter by the square root of the number $N_e$ of 
effectively independent time intervals and $N_e$ has been in turn determined as the ratio 
between the total length of the time series by the decay time of the auto-correlation function.
A long correlation time for $g=1$ ($\approx 250$ units) causes the large error compared 
to smaller ($g=0.8$) and larger coupling ($g=1.3$).

The simulations performed for three different system sizes ($4000$, $16000$ and $64000$ units) reveal the existence 
of a critical $g$-value above which $R$ grows from zero,
as in the usual Kuramoto setup and that this value is in good agreement with $g_c$ as estimated
from the linear stability analysis discussed in the previous section.
At variance with the Kuramoto model, here, the standard deviation $\sigma_R$ is larger than zero, 
meaning that the degree of synchronization oscillates in time already slightly above threshold.
Finally, $\sigma_Y$ exhibits the same behavior as $\sigma_R$, confirming that the transition is accompanied 
by the onset of macroscopic oscillations.

Finally, the red curve tracks the mean-field obtained by decreasing $g$. The difference observed
in the critical region with respect to the previous curves (obtained by increasing $g$) suggests the
possible co-existence of an asynchronous with a partially synchronised regime. Since, however,
no jump is observed in the simulations performed by increasing $g$, it is reasonable to conclude
that the bifurcation is ``supercritical" and thus to attribute such deviations to the finite
sweeping time. Anyway, since the main  goal of this work is to characterise the behavior above threshold,
we have preferred to focus our efforts on larger $g$-values, where the
asymptotic regime is much less dependent on the selection of the initial condition.

\begin{figure}[htb]
\begin{center}
\includegraphics[width=0.9\columnwidth,clip=true]{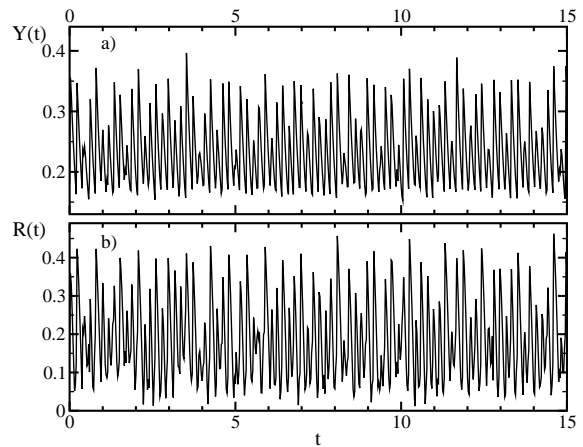}
\caption{Time evaluation of $Y$ (a) and R (b) for $g=1.3$ and system sizes: $N=4000$.}
\label{fig:timeseries}
\end{center}
\end{figure}

A first qualitative instance of the collective dynamics can be appreciated in Fig.~\ref{fig:timeseries},
where $R(t) $ and $Y(t)$ are plotted for $g=1.3$ showing that the evolution is more complex than just periodic.

\begin{figure}[htb]
\begin{center}
\includegraphics[width=0.9\columnwidth,clip=true]{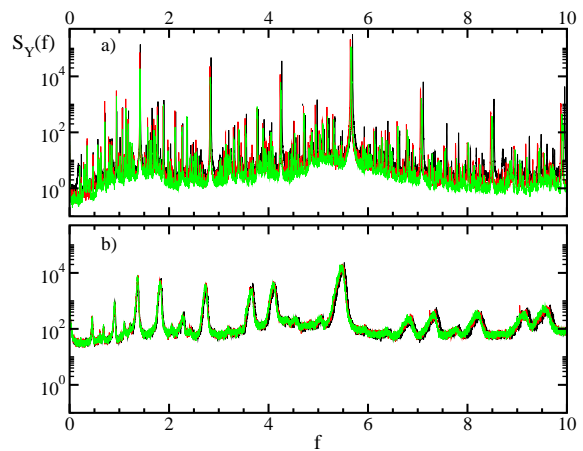}
\caption{Power spectra of $Y$ for $g=0.8$ (a), and 1.3 
(b) in each case for: $N=4000$ (black), $N=16000$ (red) and $N=64000$ (green). The
spectra are obtained by transforming time series of $819.175$ time units, 
sampled every $0.025$ units and averaged over 
$50$ different realizations.
}
\label{fig:pspect}
\end{center}
\end{figure}

A more quantitative characterization of the collective temporal behavior can be obtained by looking at
power spectra. The square amplitude of the Fourier transform of $Y(t)$ is reported in
Fig.~\ref{fig:pspect} for two different coupling strengths.

There we see that the spectra possess quite a rich structure,
being neither trivially broad-band, nor just revealing a periodic behavior (especially for $g=0.8$).
A closer look at the width of the various peaks upon increasing the network size
reveals that they do not decrease. Simulations performed for different
realization of the bare frequencies (data not shown) indicate that the results are almost independent, especially
for the larger system sizes. Altogether, we are thus led to conclude that the stochastic-like dynamics is
not due to finite-size effects, but intrinsic of the thermodynamic limit.

According to the theory of nonlinear dynamical systems, it is well known that an irregular evolution
may well be the manifestation of low-dimensional deterministic chaos. Can it be the case here?
In order to clarify the point, it is natural to investigate the behavior of the activity field 
by performing a nonlinear time series analysis, to determine its fractal dimension. 
Given a time series $Y(t_n)$, sampled
at equally spaced times ($\Delta t= t_{n+1}-t_n= 0.025$), one starts embedding the series 
into a space of dimension $m$, by building vectors of the type 
$[Y(t_n),Y(t_{n+s}),\ldots,Y(t_{n+(m-1)s})]$, where $s$ is suitably selected.
As often done, we have chosen $s$, so that $s\Delta t$ is close to the first minimum of the 
autocorrelation of $Y(t)$ ($s=5$, in our case).

The fractal dimension has then been estimated by using the nearest-neighbour method~\cite{Badii-Politi-85},
as it suffers of less fluctuations in the region of small distances. Given a generic time series,
$N_r$ reference points are randomly selected ($N_r = 10^5$ in our case). Each of them is compared with
an increasing number $n$ of randomly selected measurement points (the other points in the time series -
up to a maximum $N_m=16\cdot 10^6$), monitoring the distance $\varepsilon_m(k,n)$ of the $k$-th neighbour 
(the distance is herein estimated using the maximum norm),
for different values of the embedding dimension $m$ and $k$. A well established theory~\cite{Badii-Politi-85},
implies that for large $n$
\[
\langle \ln \varepsilon_m(k,n)\rangle \approx \frac{\ln n}{D_e} \; ,
\]
where the angular brackets denote the average over the reference points, while 
$D_e$ is the (effective) information dimension. 
In order to make the dependence of $D_e$ on the resolution $\varepsilon_m$ transparent, 
we have modified the standard approach. Once interpreted the logarithmic derivative of $n$
as a resolution-dependent dimension,
\[
D_e(\varepsilon_m) =  \frac{d \ln n}{d \langle \ln \varepsilon_m\rangle } \; ,
\]
we have plotted it versus $\langle \varepsilon_m\rangle$ itself, interpreted as an independent variable. In fact, we have verified that $D_e(\varepsilon_m)$ takes the same value, 
irrespective of the way $\varepsilon_m$ has been determined (i.e. independently of the $k$ value). 
The only differences are that larger $k$-values yield smaller statistical fluctuations, 
but are confined to larger distances. 
A good compromise has been obtained by gluing together the data obtained for the largest $k$ value (30) with the
data obtained for the smallest distances and a lower-order neighbour (4th one).

\begin{figure}[htb]
\begin{center}
\includegraphics[width=0.49\columnwidth,clip=true]{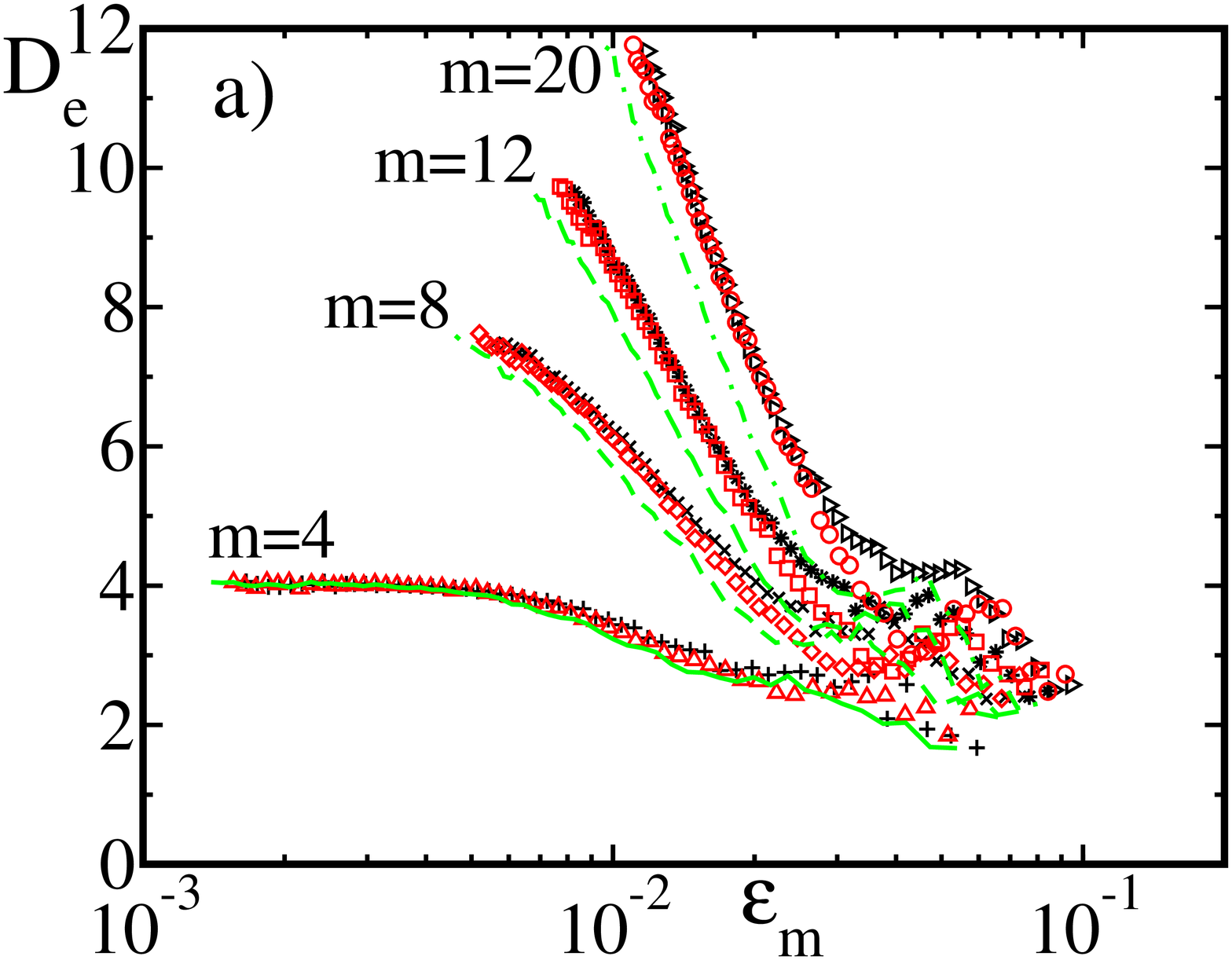}
\includegraphics[width=0.49\columnwidth,clip=true]{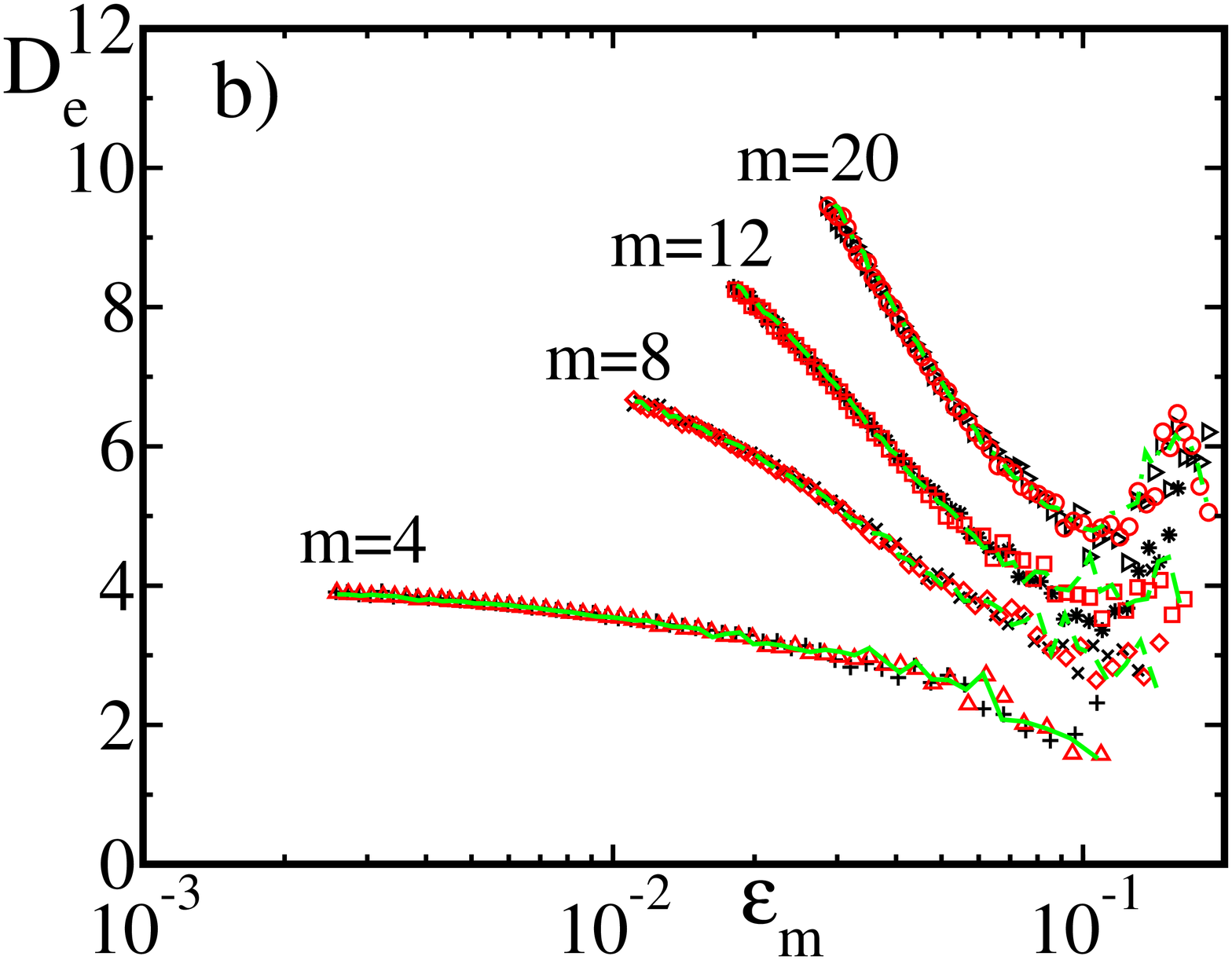}
\caption{Effective dimension $D_e$ as a function of the resolution $\varepsilon_m$ for $g=0.8$ and $g=1.3$ in panel a and b, respectively. The system size is $N=4000$ (black), $N=16000$ (red) and $N=64000$ (green). The different symbols belong to different embedding dimensions $m$ marked in the figure. The curves for the same embedding dimension group together with a similar slope irrespective of the system size.}
\label{fig:dim}
\end{center}
\end{figure}
The results for $g=0.8$ are reported in Fig.~\ref{fig:dim}a and different ensemble sizes
($N=4000$, $16000$, and $64000$). The first point to notice is that for the lowest
embedding dimension ($m=4$), $D_e$ nicely converges to 4 upon decreasing
$\varepsilon$. This clearly implies that the dimension of the collective motion is at least 4,
i.e. one needs at least four variables to characterize such a behavior. 
Furthermore, the curves obtained for the larger $m$ values, reveal an increase, each possibly
hinting to $m$, thus suggesting that the dynamics is high-dimensional (if not even
infinite dimensional). Additionally, one can also appreciate a small shift to smaller
scales of the curves obtained for $N=64000$. 
In itself, this is the indication of finite-size effects. If the shift continues as such by further 
increasing $m$, it would mean that part of the high-dimensionality is just a consequence of
statistical fluctuations which disappear in the thermodynamic limit. 
We are more inclined to attribute such discrepancies to another type of finite-size effect:
a non perfect equivalence among the various realizations of the frequency distributions.
We have, in fact, observed that different clusters may temporarily form during the evolution, especially
in the interval $g \in [g_c,1.2]$ (see a more detailed discussion in section \ref{sec:micro}).

For $g=1.3$ the convergence to the thermodynamic limit is more clear. In Fig.~\ref{fig:dim}b
the agreement among the different network sizes is compelling over a wide range, suggesting
thus that the statistical fluctuations do not affect the dimension estimates. 
We have, also, double-checked the results by computing the correlation dimension 
with the TISEAN package \cite{TISEAN-99}: a rather similar pattern emerges (data not shown). 

Finally, we have investigated the degree of (in)stability of the dynamics, computing the first 10 Lyapunov exponents $\Lambda_j$. We have followed the approach described in \cite{Olmi-etal-10}, which consists
in formally interpreting the time evolution as a series of discrete-time maps from one to the next spike emission.
The results are plotted in Fig.~\ref{fig:phasediag}c upon varying the coupling strength. 
There we see that the dynamics is always stable (notice that the zero Lyapunov exponent,
always present in a non-constant autonomous dynamics is automatically discarded).
For a vanishing $g$, all the Lyapunov exponents converge to zero: this is obvious, since in this limit
all the oscillators are uncoupled.
Much less trivial is that the Lyapunov exponents are all negatives in spite of a dynamics that may even
be collectively irregular. This manifestation of stable chaos strongly suggests that the 
connection between different levels of descriptions (micro vs. macroscopic) of a given model is weak,
if any.

\section{Microscopic dynamics}
\label{sec:micro}

\begin{figure}[htb]
\begin{center}
\includegraphics[width=0.45\textwidth,clip=true]{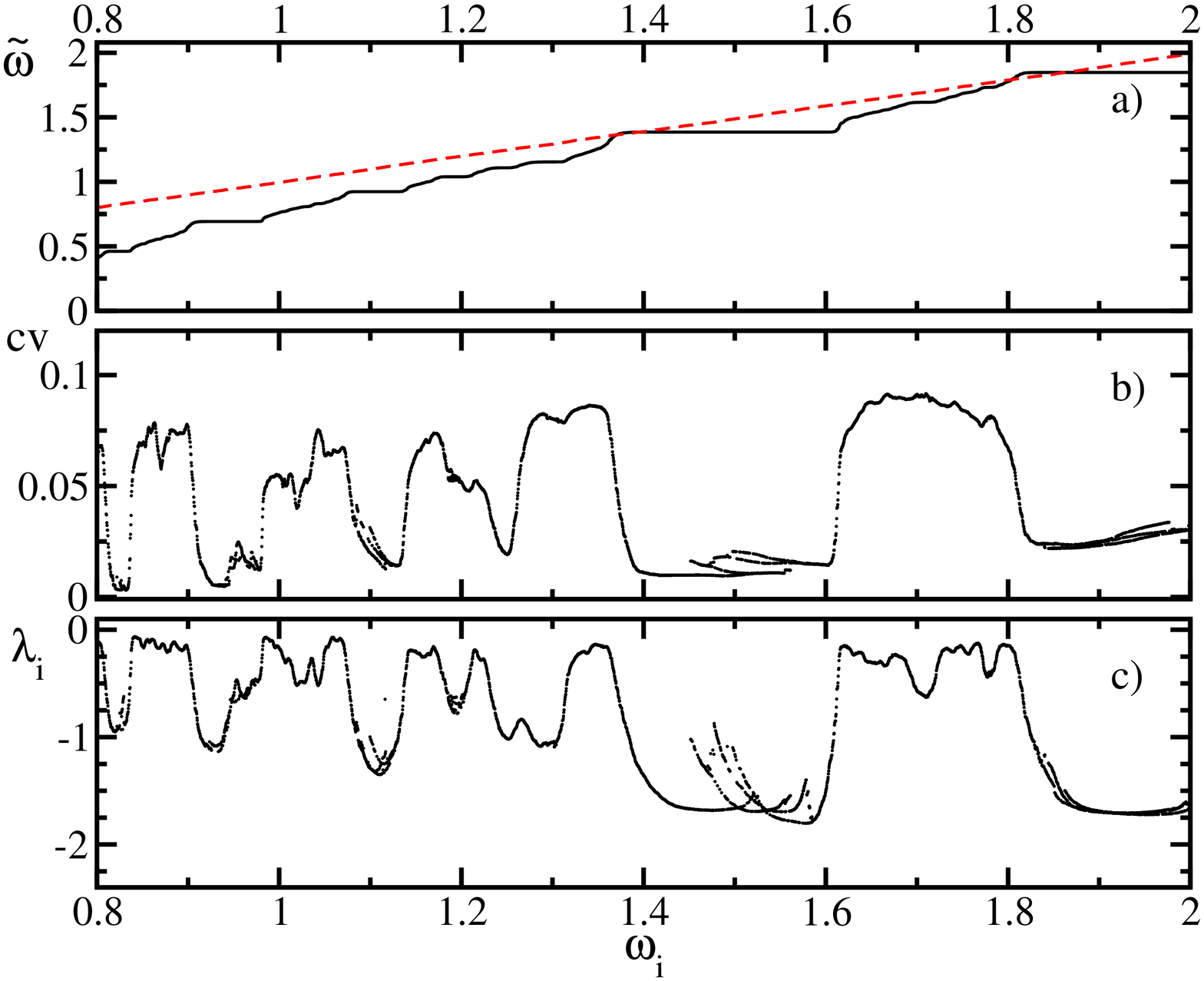}
\includegraphics[width=0.45\textwidth,clip=true]{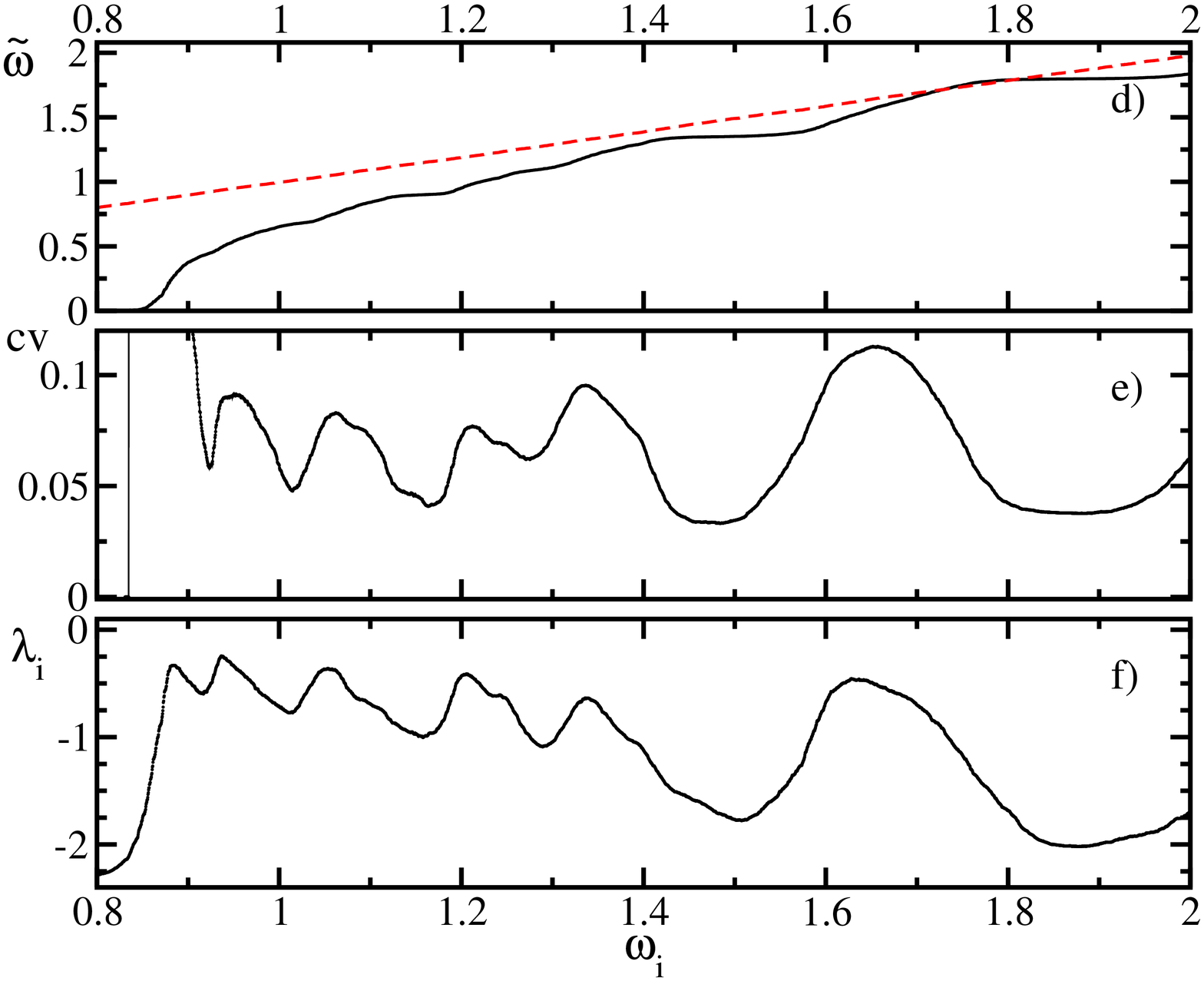}
\caption{Effective frequency $\tilde \omega$ (panels a and d), CV of $\tilde \omega$ (panels b and e) and conditional Lyapunov exponents $\lambda_i$ (panels c and f) for the different oscillators, for $g=1$ (panels a-c) and $g=1.3$ (panels d-f), all for $N=4000$. The red dashed line in panels a and d show the bare frequency as a reference for the effective frequency shown with solid black lines.}
\label{fig:period}
\end{center}
\end{figure}

In this section we try to shed light on the collective dynamics by analysing the
behavior of the single neurons. We start noticing that the coupling modifies the firing rate of
the neurons. This can be appreciated in Fig.~\ref{fig:period}a, where the effective (average) 
frequency $\tilde \omega$ is reported for the coupling strength $g=1$. The dashed line corresponds 
to the bare frequency of each neuron.
Almost everywhere $\tilde \omega$ is smaller than the bare frequency $\omega$. This is a consequence
of the fact that, although the PRC was chosen to be symmetric around zero, this is no longer true for the
effective PRC (see Fig.~\ref{fig:PRC}). The most interesting feature to notice is, however, the
staircase structure of $\tilde \omega(\omega)$ with flat plateaus which correspond to clusters 
of mutually synchronised neurons: 
the synchronization does not mean a perfect phase locking but that the phase differences 
never become larger than $2\pi$. 

One way to characterize the irregularity of the single-neuron activity is through its coefficient of variation 
(CV), i.e. the standard deviation of the the inter-spike interval rescaled to its average value.
In Fig.~\ref{fig:period}b, we see that the CV allows identifying synchronized clusters 
as those frequency intervals where the fluctuations are significantly smaller.
Furthermore, distinct lines can be recognized inside some clusters: they correspond to
different locked states \footnote{As stated previously, they are not truly phase-locked since
the phase difference fluctuates.} and are a manifestation of the multistability that
is in fact seen also at the macroscopic level.
On a more quantitative level, the neural dynamics is not significantly irregular if compared,
for instance, to the true brain activity in the resting state. It should be, however, kept in
mind that in our toy model, the only source of disorder is the distribution of bare
frequencies; no disorder has been assumed in the synaptic connections.

Additional information can be extracted by assuming that the self-determined activity field
$E(t)$ is externally given, so that each neuron can be interpreted as a forced dynamical
system. In this way, it is natural to compute the (conditional) Lyapunov exponent $\lambda_i$.
In Fig.~\ref{fig:PRC}c, one can observe a scenario that is qualitatively different from
what observed in the Kuramoto model: all $\lambda_i$'s are negative, including those
of the neurons outside the flat plateaus. We come back to this point ahead in this section.

A partially different scenario is found for $g=1.3$ (see Fig.~\ref{fig:period}d-f). First of all any sign of 
multistability has disappeared and all plateaus as well. 
The initial high peak of the CV (panel e) is due to the fact that now 
the neurons with the lowest bare frequency do not spike at all, having undergone a kind of oscillation-death. 
Their CV is obviously equal to zero. 
As a result, the first erraticaly spiking neurons are characterized by long interspike intervals 
and are obviously accompanied by large fluctuations. The CV of these rarely firing neurons is as
large as 1.75 (not shown in Fig.~\ref{fig:period}e with the present scale).
The kind of oscillation-death phenomena could be seen as an inhomogeneous limit cycle (IHLC) 
because non spiking neurons are not trapped at a steady state, rather moving back and 
forth according to their bare frequency and the global pulses,
never reaching the threshold \cite{Ullner-prl-07,*Koseska-physr-13}. 
Hence we observe two groups of neurons, the quiet neurons but still being sub-thresholdly 
active and the firing neurons. The situation is a more complicated than usual IHLC, because of the many 
(in the thermodynamic limit infinite many) frequencies reflected in the power spectra (Fig.~\ref{fig:pspect}). 
Moreover, the dynamics of each oscillator is rather stable.

Altogether, the microscopic analysis confirms that the microscopic behavior is linearly stable:
each neuron is synchronized with the self-generated mean-field $E(t)$ and yet an irregular
dynamics is self-sustained. This can be understood by noticing that in many frequency
ranges, the effective frequency is a strictly monotonous function of $\omega$.
This means that, even though each neuron synchronizes with the field $E$, the
parameter (bare frequency) mismatch induces a qualitatively different response:
such qualitative differences are then responsible for the maintenance of the
self-generated irregularity. In a more technical way: the response of a phase
oscillator is not structurally stable: one can slightly modify its frequency
and still observe significant changes (phase-slips).
A more physical (although still qualitative)  way to understand the phenomenon
is as follows: one can see each phase-oscillator as a particle moving in a potential 
with an inclination that depends on its bare frequency and the mean-field $E$.
When two particles with slightly different $\omega$ are followed, it may happen that
one of them is blocked in a (shallow) minimum which is absent for the other.

One can learn a bit more about the dependence on $\omega$  by comparing pairs of consecutive
oscillators (consecutive in the space of bare frequencies). This can be done, by monitoring phase slips,
i.e. the time instants when the phase difference becomes larger (smaller) than
$1/2$ ($-1/2$). 
Since it is possible that the phase difference may oscillate around $1/2$ ($-1/2$) yielding 
long sequences of positive and negative slips, we have chosen to record only those events
where two or more consecutive positive (negative) flips are observed.

\begin{figure}[htb]
\begin{center}
\includegraphics[width=0.49\columnwidth,clip=true]{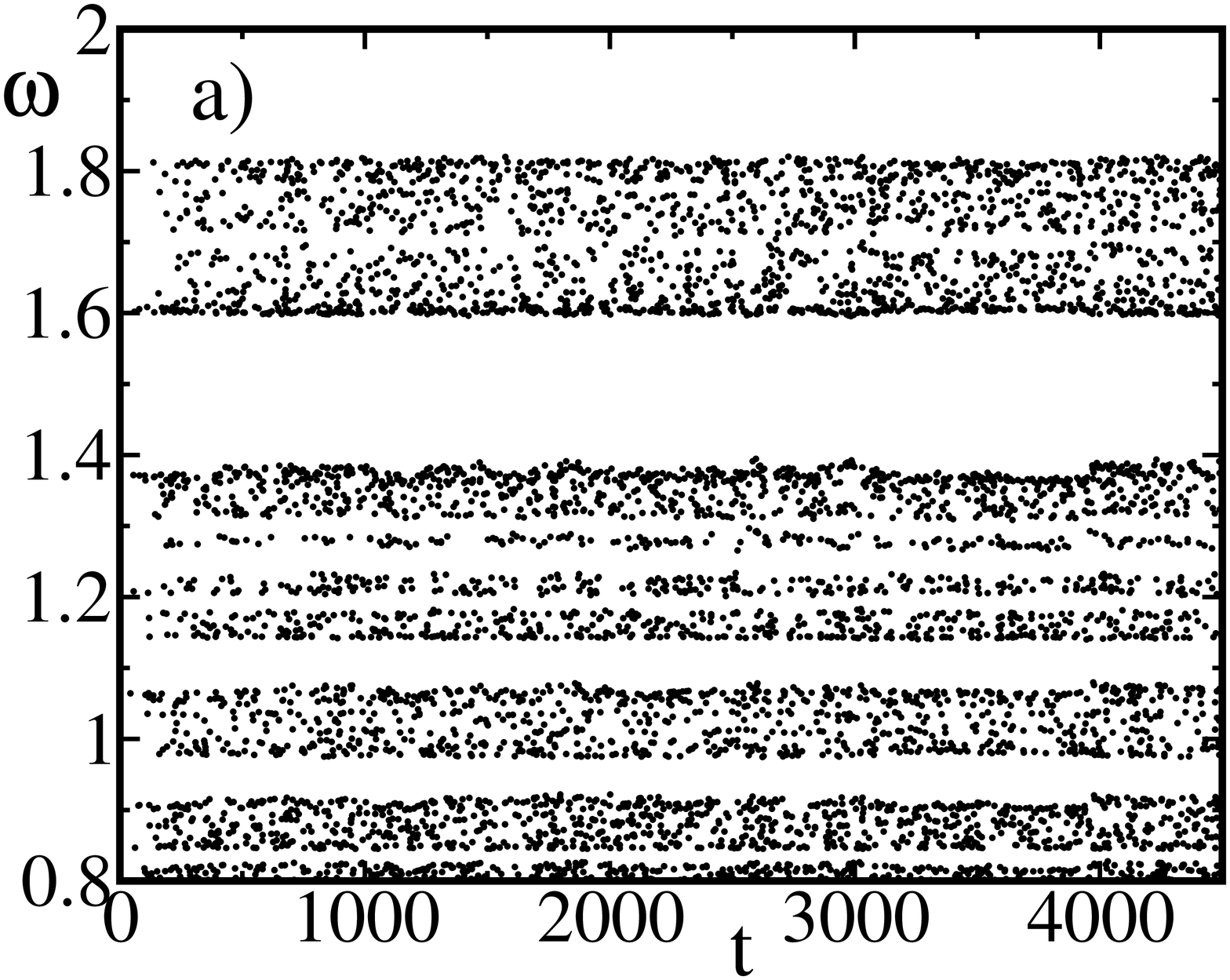}
\includegraphics[width=0.49\columnwidth,clip=true]{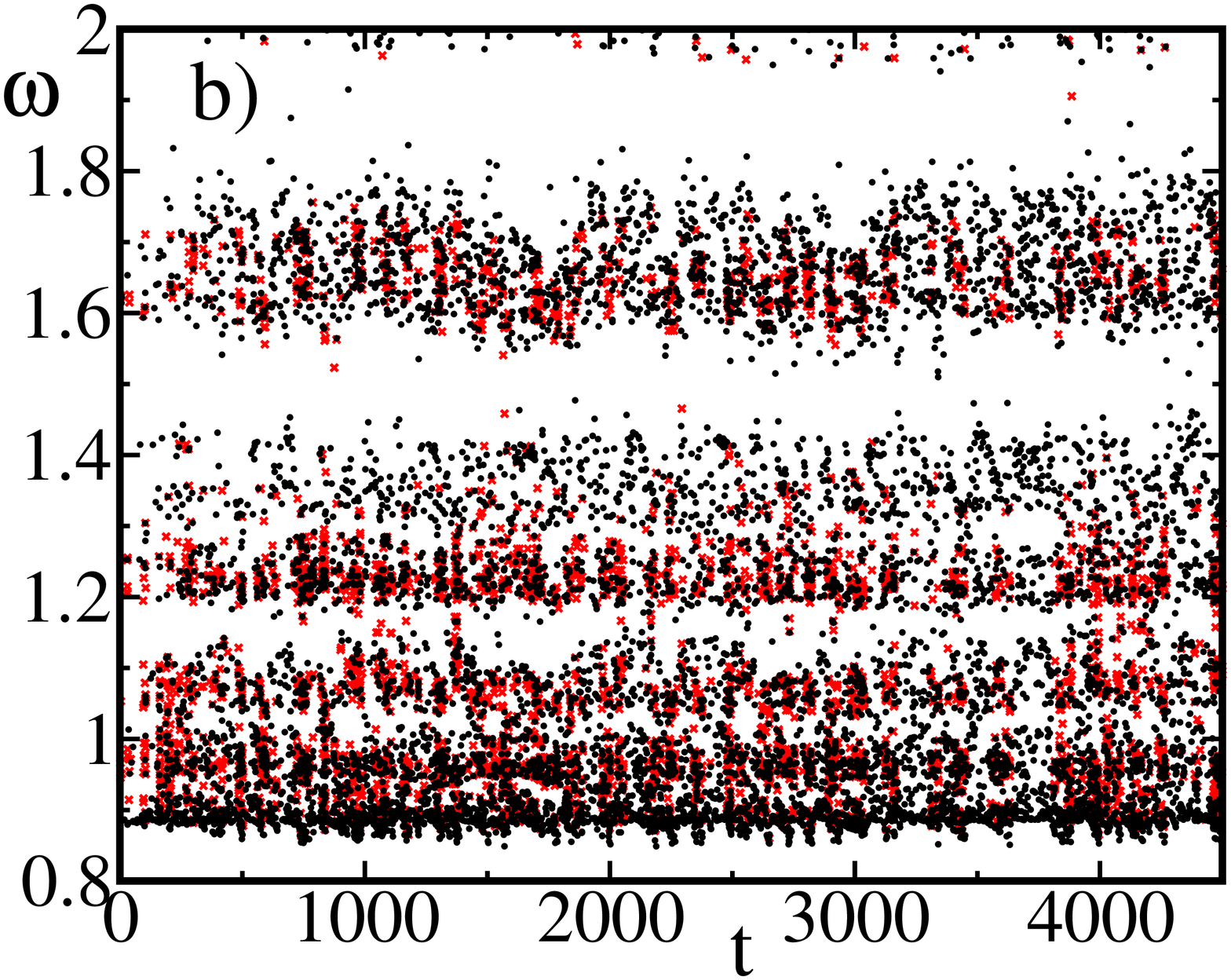}
\caption{Phase slips for $g=1$ (panel a) and $g=1.3$ (panel b) for $N=4000$.
Black circles correspond to forward slips; red crosses to backward slips, occurring only in panel b.}
\label{fig:pslips}
\end{center}
\end{figure}

In Fig.~\ref{fig:pslips}a we see that for $g=1$ there exist totally empty bands: they correspond
to the previously mentioned synchronization areas. 
In panel b (which corresponds to $g=1.3$) both forward
and backward slips are simultaneously present. One can see that the phase slips happen 
on a much longer time scale than the interspike intervals. 
Further, the bands with sparse phase slips observed for $g=1.3$ are reminiscent of the synchronization
bands found for $g=1.0$: there,  phase slip events are rare and erratic. 
The totally empty band at the bottom frequencies for $g=1.3$ corresponds to the non-firing neurons.

\begin{figure}[htb]
\begin{center}
\includegraphics[width=0.49\columnwidth,clip=true]{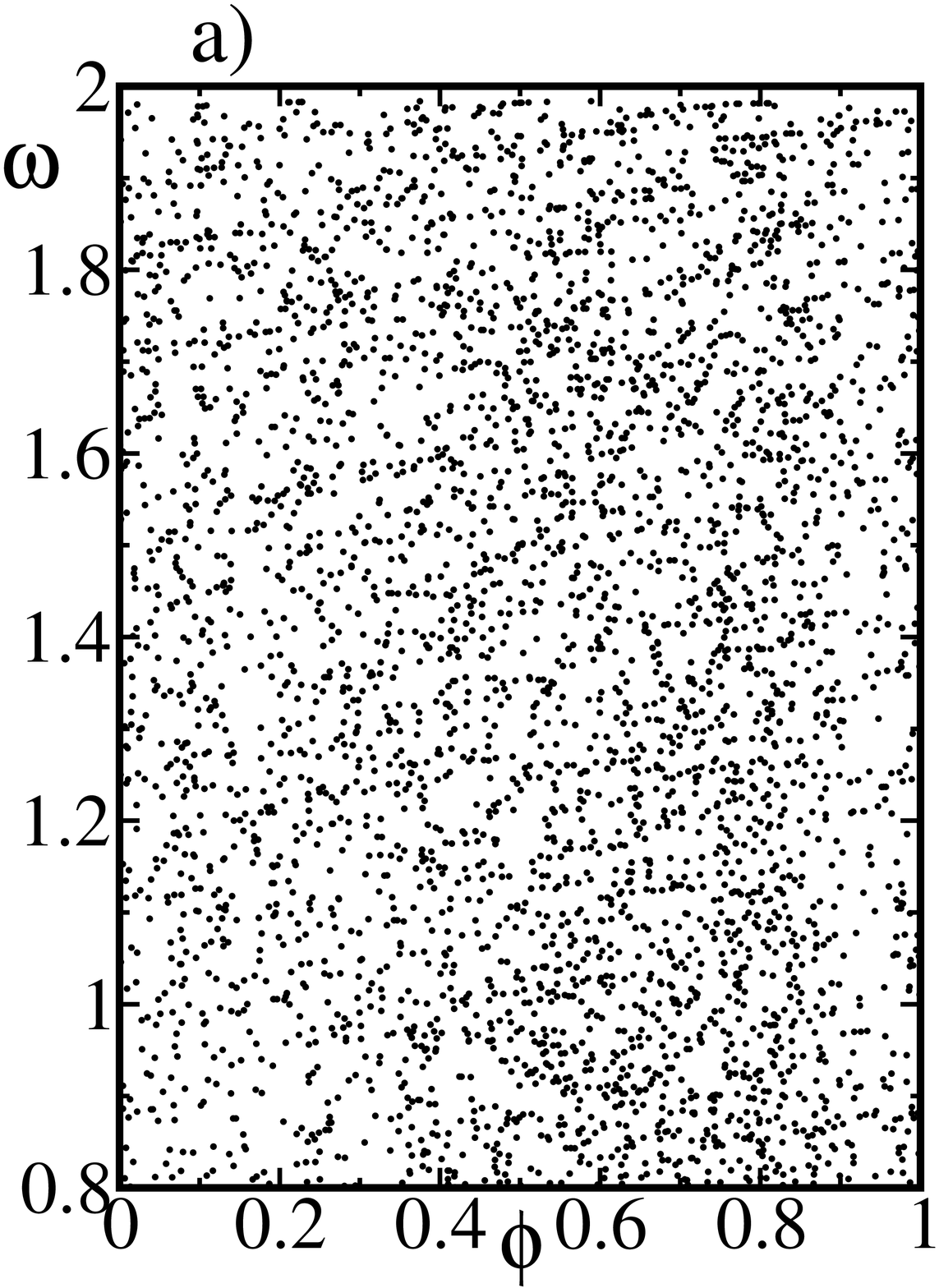}
\includegraphics[width=0.49\columnwidth,clip=true]{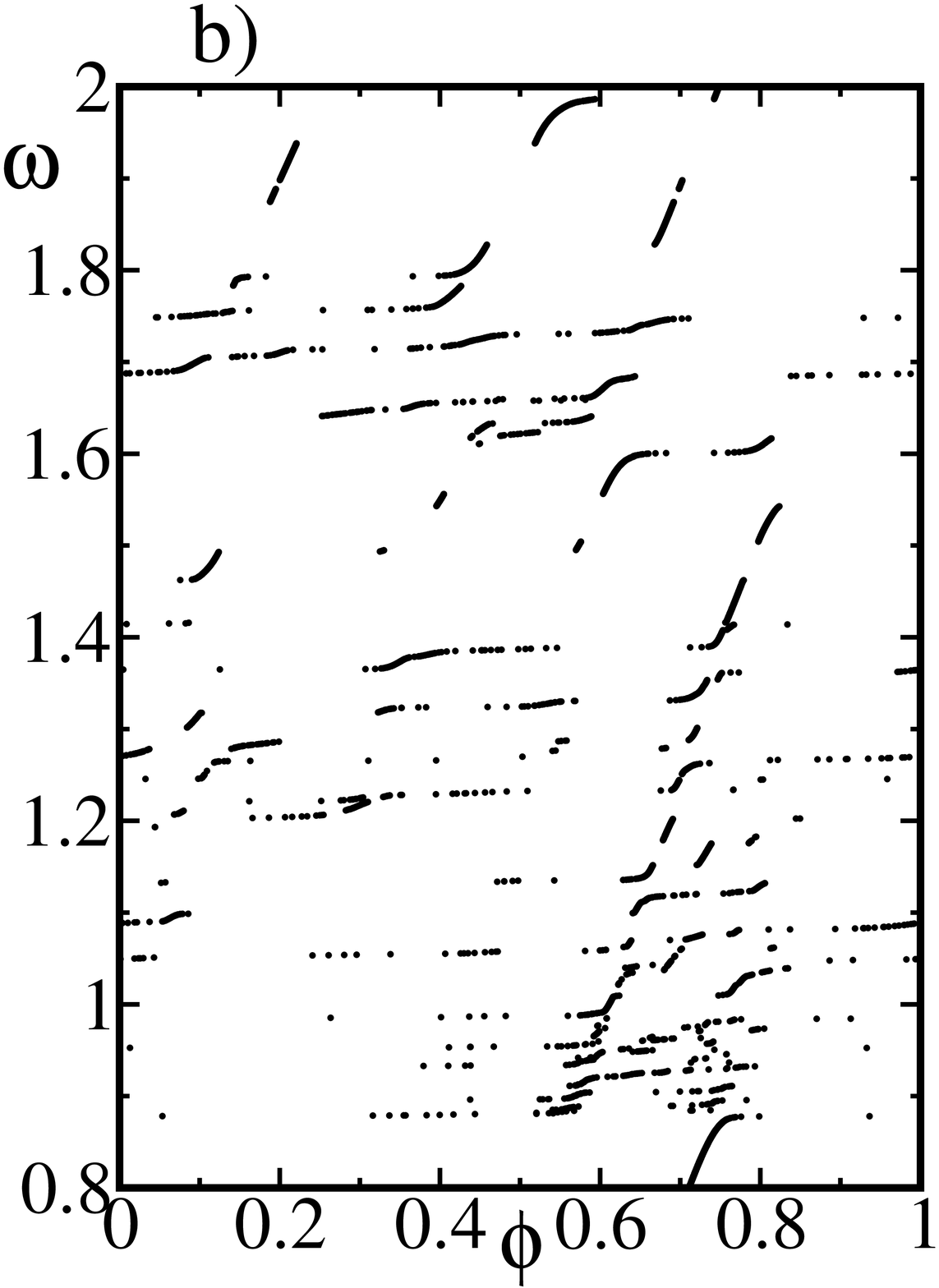}
\caption{Snapshots of the probability density $Q(\phi,\omega,t)$ for $g=0.5$ 
(panel a) and $g=1.3$ (panel b) and $N=4000$.}
\label{fig:snapshot}
\end{center}
\end{figure}

In the thermodynamic limit, the most appropriate way to characterize the collective dynamics is
by monitoring the probability distribution $Q(\phi,\omega,t)$ introduced in Sec.~\ref{sec:theory}.
In Fig.~\ref{fig:snapshot} we give an idea of the way it looks like below and above threshold
at some randomly chosen time. In panel a) one can recognize a reasonably smooth distribution.
In fact, for $g=0.5$, we are in the asynchronous regime and thereby expect a smooth distribution
of the phases $\phi$~\footnote{A perfect uniformity would be observed if the effective phases
$\vartheta$ were to be used - see the appendix B for their definition.} Such a distribution looses
stability above $g_c$.

In fact, for $g=1.3$ we see a rather different structure (see panel b) characterized by an alternation 
of highly dense and widely spread regions. It is clear that even the plain integration of the
equations (\ref{eq:continuity},\ref{eq:field1}) is a highly nontrivial task, not to speak of
the development of approximate analytical schemes.

\section{Discussion and open problems}

In this paper we have analysed an ensemble of pulse-coupled oscillators characterized
by a distribution of bare frequencies and coupled through a homogeneous mean field.
Although the setup is reminiscent of the Kuramoto model, the collective dynamics is much richer
and accompanied by a linearly stable microscopic dynamics.

A linear stability analysis of the asynchronous regime allows identifying the transition 
point beyond which a complex form of synchronization sets in. A numerical analysis of
a properly defined Kuramoto order parameter $R$ and of the smoothed activity
field $Y$ reveals that they not only fluctuate in time, but their behavior involves
a large (possibly infinite) number of degrees of freedom.
This indicates that even in ``simple" mean-field models, such as the one investigated
in this paper, the coarse-grained activity of an ensemble of phase-oscillators
cannot be reduced to the evolution of one or a few variables, such as the firing rate
and related observables. In principle, nothing prevents a population of phase-oscillators
to self-sustain a macroscopic irregular dynamics: the corresponding evolution equation
is indeed a nonlinear functional equation (see Eqs.~(\ref{eq:continuity},\ref{eq:field1})),
which operates in an infinite-dimensional phase-space.
It is, however, unclear under which conditions many degrees of freedom can be simultaneously
active. In order to make further progress, it will be necessary to
find suitable approximations of the probability density $Q(\phi,\omega,t)$: 
this task seems to require clever ideas on the way to expand $Q(\phi,\omega,t)$.
One question is particularly relevant: whether the dynamics is born high
dimensional from the very beginning 
(such as in models of balanced states ~\cite{Sompolinsky-Crisanti-88,Curato-Politi-13})
or the complexity increases by undergoing a series of consecutive bifurcations. 
The numerical analysis in
the vicinity of the critical point is affected by too strong finite-size corrections
to be able to draw any conclusion.

Another open point is the generality of this scenario. Several preliminary simulations
performed with various choices of the PRC reveal that it is quite robust,
although the presence of a relatively steep branch seems to be a necessary condition.
This is not too serious a limitation, as it naturally appears in systems 
characterized by a slow-fast dynamics (see the discussion in~\cite{Langfield-etal-15}).
It might be, however, worth to assume a different PRC shape to enable deeper analytical
studies. We have indeed derived a very general equation for the loss of stability
of the asynchronous state: if one could go beyond, including the most relevant
nonlinear terms, it should be possible to decide how many degrees of freedom
are switched on. 

Our numerical studies suggest that the transition disappears
when the distribution of frequencies is narrow enough, but this is by no means a
proof: understanding whether it is strictly necessary to go beyond the weak-coupling,
weak-disorder limit is another point that will be worth exploring. 

Another intriguing property of the collective dynamics discussed in this paper is
the presence of a spectrum of negative Lyapunov exponents. 
This means that it is a manifestation of {\it stable chaos} \cite{Politi-Torcini-10}. 
Within the context of computational neuroscience, the stability of the microscopic trajectories
suggests that this model is a good candidate for performing computational tasks. 
It will be worth to explore this opportunity by studying the response of this type of networks 
to different classes of external stimuli.

\section*{Acknowledgment}
AP wishes to acknowledge a discussion with M. Wolfrum on the stability of the Kuramoto model.

\appendix
\section{Stability analysis}
The equation (\ref{eq:afterA}) for $u$ can be rewritten as
\[
\frac{du}{d\phi} =
\frac{(g\Gamma'(\phi) E_0 -\mu )u + g [\Gamma'(\phi)Q_0 
+ \Gamma(\phi)\frac{dQ_0}{d\phi}] z}
{\omega - g \Gamma(\phi) E_0 }
\]
which has the structure
\begin{equation}
\frac{du}{d\phi} = (A-\mu \tau) u + g P(\omega) C(\phi,\omega)z
\label{eq:dudphi}
\end{equation}
where
\begin{equation}
A = \frac{g\Gamma'(\phi) E_0} {\omega - g \Gamma(\phi) E_0 } \quad ,
\label{eq:A01}
\end{equation}
\begin{equation}
C(\phi,\omega) = 
\frac{\omega \Gamma'(\phi)} {T(\omega,E_0)(\omega - g \Gamma(\phi) E_0)^3 } 
\label{eq:B}
\end{equation}
while $\tau$ is defined in (\ref{eq:Tomega}).
The general solution of  Eq.~(\ref{eq:dudphi}) is
\begin{eqnarray}
u(\phi,\omega) &=&  \textrm{e}^{F(\phi,\omega)-\mu T(\phi,\omega)} \left [ u(0,\omega) + \right .
\label{eq:uphi} \\
&& \left . 
z \int_0^\phi d \psi C(\psi,\omega) \textrm{e}^{-F(\psi,\omega)+\mu T(\psi,\omega)}\right ]
\nonumber
\end{eqnarray}
where
\begin{equation}
F(\psi,\omega) = \int_0^\psi d\eta A(\eta) =
\log \frac{\omega-g\Gamma(0)E_0}{\omega-g\Gamma(\psi)E_0} \quad ,
\label{eq:FG}
\end{equation}
notice that $F(1)=F(0)=0$, 
while $T(\psi,\omega)$ is defined in Eq.~(\ref{eq:Tomega}).
One can therefore write the solution $u(\phi,\omega)$ as
\begin{eqnarray}
u(\phi,\omega) &=& \frac{\omega-g\Gamma(0)E_0}{\omega-g\Gamma(\phi)E_0} 
\textrm{e}^{-\mu T(\phi,\omega)} \left [ u(0,\omega) + \right . \nonumber \\
&& \left .
z\, g\, P(\omega) \frac{V_\mu(\phi,\omega)}{\omega-g \Gamma(0)E_0}\right ]
\label{eq:uphi2}
\end{eqnarray}
where
\begin{equation}
V_\mu(\phi,\omega) = \frac{\omega}{T(1,\omega)}
\int_0^\phi d \psi
\frac{\Gamma'(\psi) \textrm{e}^{\mu(\psi,\omega)}} {(\omega - g \Gamma(\psi) E_0)^2} 
\end{equation}
By imposing the periodicity condition $u(1,\omega)=u(0,\omega)$ one obtains
\begin{equation}
u(1,\omega) = g\, z 
\frac{P(\omega) V_\mu(1,\omega)(1,\omega)} {(\textrm{e}^{\mu T(1,\omega)}-1)(\omega-g\Gamma(0)E_0)} 
\label{eq:ueps}
\end{equation}
By finally, recalling the definition (\ref{eq:epsdef}) $z$, it is possible 
to rewrite Eq.~(\ref{eq:ueps}) as the eigenvalue equation
\begin{equation}
1+g\Gamma(1)\int Q_0(1,\omega)d \omega =
g \int d\omega \frac{P(\omega)V_\mu(1,\omega)} {\textrm{e}^{\mu T(1,\omega)}-1}
\label{eq:mudef}
\end{equation}

\section{Transformation to the appropriate Kuramoto like phase $\vartheta$}

We solve the original system Eq.~(\ref{eq:pulsecoup}) with the PRC Eq.~(\ref{eq:PRC}) in terms of the non-homogeneously advancing phase $\phi$. At each time point we calculate the Kuramoto order parameter $R$ we convert the phase $\phi$ into the Kuramoto like phase $\vartheta$ according to Eqs.~(\ref{eq:phase2}) and (\ref{eq:effPRC}). For simplicity and readability we introduce new constants similar those defining the piecewise linear PRC Eq.~(\ref{eq:PRC}): $\mathfrak{B}_{01}=\frac{g}{N} E_0 B_{01}$, $\mathfrak{B}_{02}=\frac{g}{N} E_0 B_{02}$, $\mathfrak{B}_{03}=\frac{g}{N} E_0 B_{03}$, $\mathfrak{b}_{1}=\frac{g}{N} E_0 b_{1}$, and $\mathfrak{b}_{2}=\frac{g}{N} E_0 b_{2}$. Hence the Kuramoto like phase is

\begin{equation}
\vartheta(\phi(t)) =
\begin{cases}
\frac{\tilde{\omega}}{\mathfrak{b}_1}  
\ln{\frac{\omega - \mathfrak{B}_{01}}
{\omega - \mathfrak{B}_{01} - \mathfrak{b}_1 \phi(t)}}
& \mbox{if } \quad 0\le\phi<\phi_l \\
\vartheta_l - 
\frac{\tilde{\omega}}{\mathfrak{b}_2} 
 \ln{\frac{\omega - \mathfrak{B}_{02} + \mathfrak{b}_2 \phi_l}
 {\omega - \mathfrak{B}_{02} + \mathfrak{b}_2 \phi(t)}}
& \mbox{if } \quad \phi_l \le \phi \le \phi_r \\
\vartheta_r + 
\frac{\tilde{\omega}}{\mathfrak{b}_1}  
\ln{\frac{\omega - \mathfrak{B}_{03} - \mathfrak{b}_1 \phi_r}
{\omega - \mathfrak{B}_{03} - \mathfrak{b}_1 \phi(t)}}
& \mbox{if } \quad \phi_r < \phi < 1 \; ,
\end{cases}
\nonumber
\end{equation}

with the field $E_0$ as stated in Eq.~(\ref{eq:E0}) and the effective frequency defined as the inverse interspike interval, i.e. $\tilde{\omega} = 1/T(1,\omega,E_0)$ (Eq.~(\ref{eq:Tomega})). The interspike interval can be expressed explicitly for the give PRC:

\begin{eqnarray}
T(1,\omega,E_0) &=&  
\frac{1}{\mathfrak{b}_1} 
\ln{\frac{\omega - \mathfrak{B}_{01}}
{\omega - \mathfrak{B}_{01} - \mathfrak{b}_1 \phi_l}}
\nonumber \\
&&  
- \frac{1}{\mathfrak{b}_2}
 \ln{\frac{\omega - \mathfrak{B}_{02} + \mathfrak{b}_2 \phi_l}
 {\omega - \mathfrak{B}_{02} + \mathfrak{b}_2 \phi_r}}
\nonumber \\
&&
+ \frac{1}{\mathfrak{b}_1} 
\ln{\frac{\omega - \mathfrak{B}_{03} - \mathfrak{b}_1 \phi_r}
{\omega - \mathfrak{B}_{03} - \mathfrak{b}_1}} 
\quad.
\nonumber
\end{eqnarray}

As shown in Fig.~\ref{fig:PRC}, the transitions $\vartheta_l$ and $\vartheta_r$ in the effective PRC $\tilde{\Gamma}$ depend on the bare frequency $\omega$ according to:

\begin{eqnarray}
\vartheta_l &=&  \frac{\tilde{\omega}}{\mathfrak{b}_1}  
\ln{\frac{\omega - \mathfrak{B}_{01}}
{\omega - \mathfrak{B}_{01} - \mathfrak{b}_1 \phi_l}}
\nonumber \\
\vartheta_r &=&  \vartheta_l - \frac{\tilde{\omega}}{\mathfrak{b}_2}  
\ln{\frac{\omega - \mathfrak{B}_{02} + \mathfrak{b}_2 \phi_l}
{\omega - \mathfrak{B}_{02} + \mathfrak{b}_2 \phi_r}}
\quad.
\nonumber
\end{eqnarray}


%

\end{document}